\documentclass[12pt]{article}
\usepackage{amssymb}
\usepackage{amsmath}
\usepackage{graphicx}
\usepackage{latexsym}

\begin{document}

\begin{center}
\vspace{1cm}{ $E_{11},K_{11}$ and $EE_{11}$}

\vspace{1cm} {\bf H. Mkrtchyan} \footnote{ E-mail:hike@r.am} and
{\bf R. Mkrtchyan} \footnote{ E-mail: mrl@r.am} \vspace{1cm}

\vspace{1cm}

{\it Theoretical Physics Department,} {\it Yerevan Physics
Institute}

{\it Alikhanian Br. St.2, Yerevan 375036, Armenia}
\end{center}

\vspace{1cm}
\begin{abstract}

In the study of conjecture on M-theory as a non-linear realization
$E_{11}/K_{11}$ we present arguments for the following: 1)roots of
$K_{11}$ coincide with the roots of Kac-Moody algebra $EE_{11}$
with Dynkin diagram given in the paper, 2)one of the fundamental
weights of $EE_{11}$ coincides with $l_1$ weight of $E_{11}$,
known to contain 11d supergravity brane charges. The statement 1)
is extended on $E_{10}$ and $E_9$ algebras.
\end{abstract}

\renewcommand{\thefootnote}{\arabic{footnote}}
\setcounter{footnote}0 {\smallskip \pagebreak }

\section{Introduction}

The search for appropriate description of superstrings/M theory
remains one of the important theoretical goals. It was conjectured
\cite{{w0},{w2},{w4},west}that nonlinear realization of $E_{11}$
Kac-Moody algebra (group) can be relevant for M-theory. $E_n,
n\leq10$ groups appeared previously as U-duality symmetries of
dimensionally reduced M-theory,  and $E_{10}$ has been suggested
as a candidate for possible description of M-theory
\cite{n4,n3,n2}. The $E_{11}$ conjecture join the $E_n$ symmetry,
usually linked to dimensionally reduced theory, and full
multidimensional dependence of (infinitely many) fields of
nonlinear realization sigma-model $E_{11}/K_{11}$. There is no
precise form of the conjecture, and some important ingredients of
the theory, e.g. description of space-time, fermionic content and
supersymmetry relations remain unclear. Connection of this
hypothesis to some other lines of search of hidden symmetries of
M-theory, particularly to $OSp(1\mid64)$ conjecture \cite{w1} is
also unclear. As mentioned in \cite{k2}, corresponding even
generators appear at low levels of $E_{11}$, but their commutators
do not close into Sp(64). In the present paper we would like to
contribute to clarification of some group-theory aspects of
$E_{11}/K_{11}$ nonlinear model, and present some observations,
which can help in inclusion of supersymmetry in the model. Our
main attention is the gauged subgroup $K_{11}$. It is described as
a subgroup of $E_{11}$, invariant w.r.t. the Chevalley involution
(cf. Section 2). It is not described as a Kac-Moody algebra, and
one of the problems we addressed to is to clarify how close is it
to some Kac-Moody algebra. The first step, in Section 2, is to
find the rank of $K_{11}$, and it was shown to be 11, as implied
by its notation. In Section 3 we suggest to consider the problem
with respect to new Cartan basis, assuming, in analogy with
finite-dimensional simple groups, that all Kac-Moody defining
relations have the same form in the new basis. We suggest a
definition of $K_{11}$ in a new basis, identify some Kac-Moody
subalgebra of $K_{11}$, which we call $EE_{11}$, with Dynkin
diagram at (\ref{ee11}) and argue that roots of $K_{11}$ coincide
with those of $EE_{11}$. This statement is not proved, we only
have been able to check that up to level 146 of group $E_{11}$. If
it is correct, then generators of these two groups, corresponding
to the real roots, completely coincide, because real roots have
multiplicity one. Difference can arise in the sector of imaginary
roots only. Similar statements are right for $E_9$ and $E_{10}$,
the Dynkin diagrams of corresponding subalgebras are given at
(\ref{d91}), (\ref{de10}). Situation for $E_9$ seems to be even
more advanced, because both $E_9$ and $D_9^{(1)}$ (analog of
$EE_{11}$) are of affine Kac-Moody type, for which multiplicities
are known to be equal to the rank of the algebra, so, conclusion
seems to be that compact subgroup of $E_9$ is Kac-Moody algebra
with Dynkin diagram given in (\ref{d91}).

It is worth mentioning that we carried out similar calculations
for a few other groups, including $E_{12}$, and didn't find
similar phenomena.

In Section 4 we consider representations of $EE_{11}$ and
$E_{11}$. We show that one of the fundamental weights of $EE_{11}$
coincides with the $l_1$ fundamental weight of $E_{11}$, which is
identified by West \cite{w3} as containing all 11d branes charges
appearing in the r.h.s. of 11d susy relation. Perhaps this
statement can be a hint for generalization of 11d susy relation on
$E_{11}$. Moreover, it may appear to be a generalization of the
observation of de Wit-Nicolai \cite{wit}, that, depending on
dimensionality, some charges in the r.h.s. of susy relation join
into representations of corresponding $E_{n}$ duality groups.

Conclusion contains discussion of present results and possible
ways of further developments.

\section{Rank of $K_{11}$}

The non-linear realization suggested in \cite{w0} has $E_{11}$ as
a (global)  Kac-Moody group, factorized over (local) group
$K_{11}$ (here and below we, following \cite{w0}, are not making
difference in notations between groups and algebras). The Lie
algebra of $E_{11}$ is a particular case of Kac-Moody \cite{kac1}
Lie algebra with basic generators $h_i, e_i, f_i$ and relations:

\begin{eqnarray} \label{km}
[h_i,e_j]&=& A_{ij} e_j  \\\label{2} [h_i,f_j]&=&-A_{ij}f_j
\\\label{3}
[e_i,f_j]&=& \delta_{ij}h_j \\ \label{4}
ad(e_i)^{(1-A_{ij})}e_j&=&0, \\
\label{5}ad(f_i)^{(1-A_{ij})}f_j&=&0
\end{eqnarray}

All other generators should be obtained from these generators by
all possible multiple commutators, factorized over relations
(\ref{km})-(\ref{5}). In these expressions $A_{ij}$ is the Cartan
matrix, which completely characterizes the algebra, and which we
assume symmetrizable, with 2 on main diagonal, and some negative
integers off diagonal, with additional requirement that if some
off-diagonal element is non-zero, then its transposed one is
non-zero, too.

Actually in the present work only symmetric Cartan matrix will
appear, with non-zero off-diagonal elements all equal (-1).

Most important for us among them is $E_{11}$ Cartan matrix, which
is given below:

\begin{eqnarray}\label{e11c}
E_{11}=\begin{array}{*{20}c}
   2&-1&0&0&0&0&0&0&0&0&0 \\
    -1&2&-1&0&0&0&0&0&0&0&0 \\
    0&-1&2&-1&0&0&0&0&0&0&0 \\
    0&0&-1&2&-1&0&0&0&0&0&0 \\
    0&0&0&-1&2&-1&0&0&0&0&0 \\
    0&0&0&0&-1&2&-1&0&0&0&0 \\
    0&0&0&0&0&-1&2&-1&0&0&0 \\
    0&0&0&0&0&0&-1&2&-1&-1&0 \\
    0&0&0&0&0&0&0&-1&2&0&0 \\
    0&0&0&0&0&0&0&-1&0&2&-1 \\
    0&0&0&0&0&0&0&0&0&-1&2
\end{array}
\end{eqnarray}

Alternatively it can be given by Dynkin diagram:

\begin{align}\label{e11}
\begin{picture}(0,100)(0,-50)
\put(-100,0){\line(1,0){180}} \put(-100,0){\circle*{10}}
\put(-80,0){\circle*{10}} \put(-60,0){\circle*{10}}
\put(-40,0){\circle*{10}} \put(-20,0){\circle*{10}}
\put(0,0){\circle*{10}} \put(20,0){\circle*{10}}
\put(40,0){\circle*{10}} \put(40,0){\line(0,1){30}}
\put(40,30){\circle*{10}}
\put(60,0){\circle*{10}}\put(80,0){\circle*{10}}
\put(-102,-15){\text{1}}\put(-82,-15){\text{2}}
\put(-62,-15){\text{3}}\put(-42,-15){\text{4}}
\put(-22,-15){\text{5}}\put(-2,-15){\text{6}}
\put(18,-15){\text{7}}\put(38,-15){\text{8}}
\put(58,-15){\text{10}}\put(78,-15){\text{11}}
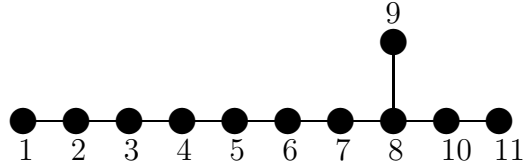
\put(37,37){\text{9}} \put(-100, -50){\text{Dynkin diagram of
$E_{11}$ algebra}}
\end{picture}
\end{align}

Subalgebra $K_{11}$  consists of generators invariant under
Chevalley involution:

\begin{eqnarray} \label{chev}
h_i\rightarrow -h_i \\
e_i\rightarrow -  f_i \\
f_i\rightarrow -e_i \\
\end{eqnarray}

These definitions evidently extend to any Kac-Moody algebra. We
shall call subalgebras $K_n$ the Chevalley-invariant subalgebras,
or compact subalgebras. As noted in \cite{w4}, \cite{k2},  some
additional signs $\epsilon_i$ should be introduced in r.h.s. of
above expressions to take into account possibility of different
signatures of space-time, and they correspond to different real
forms of $K_{11}$. Below we shall be interested in problems, which
are independent from the field of coefficients, namely in the rank
of $K_n$ and its possible Dynkin diagram, so we shall assume the
field of coefficients to be the complex numbers. We assume that
all our results will have some counterpart for different real
forms of $E_n$ and $K_n$, corresponding to different
compactifications of 11d theory (see \cite{k2, k1,k3}). As is
well-known, in a finite-dimensional case real forms all are based
on the same Dynkin diagrams, decorated to become Vogan or Satake
diagrams \cite{knapp,helg}, so by expecting something similar for
Kac-Moody algebras, we ascertain the importance of understanding
of the relations between $K_n$ and Kac-Moody algebras with
definite Dynkin diagrams.

For a finite-dimensional $E_n$ groups there are following groups
$K_n$ (table from \cite{w0}, see also \cite{wit}):

\begin{eqnarray}
\begin{array}{*{20}c}
   \begin{array}{l}
 D \\
 11 \\
 \end{array} & \begin{array}{l}
 E_n  \\
 1 \\
 \end{array} & \begin{array}{l}
 K_n  \\
 1 \\
 \end{array}  \\
   {10,IIB} & {SL(2)} & {SO(2)}  \\
   {10,IIA} & {SO(1,1)/Z_2 } & 1  \\
   9 & {GL(2)} & {SO(2)}  \\
   8 & {E_3 \sim SL(5) \times SL(2)} & {U(2)}  \\
   7 & {E_4 \sim SL(5)} & {USp(4)}  \\
   6 & {E_5 \sim SO(5,5)} & {USp(4) \times USp(4)}  \\
   5 & {E_6 } & {USp(8)}  \\
   4 & {E_7 } & {SU(8)}  \\
   3 & {E_8 } & {SO(16)}  \\
\end{array}
\end{eqnarray}

We see that the rank of group K can differ from (be less than)
that of E. We choose the following way of finding the rank of
$K_n$. Consider the subgroup of $E_n$, generated by Dynkin diagram
of $E_n$, taken without rightmost node (such subgroup for $E_{10}$
was recently discussed in \cite{n5}). That is a Dynkin diagram of
SO(2n-2) group. We shall use the fact that Chevalley-invariant
subgroup of SO(2n) is SO(n) x SO(n). For $E_{11}$ this means that
we can find 10 Cartan generators of SO(20) which belong to its
Chevalley-invariant subgroup SO(10) x SO(10). It remains to find
one additional invariant generator, commuting with these 10 to
prove that rank of Chevalley-invariant subgroup of $E_{11}$ is 11.
With that purpose we decompose the $E_{11}$ algebra w.r.t. the
SO(20). Decomposition is naturally graded by the powers of
generators $e_{11}$ and $f_{11}$. Remind that the space of
generators consists of positive and negative roots generators,
constructed by powers (commutators) of solely e and f generators,
respectively. From relations (\ref{km})-(\ref{5}) one sees that
subspace with given power of $e_{11}$ and $f_{11}$ is invariant
w.r.t. the SO(20). So, if we find a generator constructed from e-s
and commuting with Cartan's of SO(10) x SO(10), we can add or
subtract (depending on overall power of e-s) a similar generator
constructed from f-s, and obtain the Chevalley-invariant
generator, commuting with first 10 Cartan generators. Since it is
anti-Hermitian, under appropriate definition of hermiticity, we
assume that its action will be semi-simple, so it can be
considered as 11-th Cartan generator.

We can change by SO(20) rotation the basis of Cartan generators
from initial one, with properties (\ref{chev}) under involution,
to new one, so that its ten Cartan generators will be those of
SO(10) x SO(10), which are involution-invariant Cartan generators,
and we can calculate the roots of $E_{11}$ in the new basis, and
look for the roots which have zeros at first 10 positions of its
weight vector. That will mean that commutator of corresponding
generator(s) with first 10 Cartan generators is zero and so we
have complete set of invariant Cartan generators. According to
\cite{helg}, zero weights appear in representations with highest
weight expressible through roots with integer coefficients. Since
in Kac-Moody algebras number of representations is infinite, it is
"very likely" that such representation exists. For
finite-dimensional algebras it may not be the case.

We use computer to generate roots of $E_{11}$ and indeed the
requested root is found on level (i.e. height in $E_{11}$) 47. It
is in 4-th rank antisymmetric tensor representation of SO(20),
which can be seen from the corresponding highest weight. The root
found, its weight, highest root of the module and the weight of
highest root are the following:

root = (1, 2, 3, 4, 5, 6, 7, 8, 4, 5, 2)

weight = (0, 0, 0, 0, 0, 0, 0, 0, 0, 0, -1)

highest root = (2, 4, 6, 8, 9, 10, 11, 12, 6, 7, 2)

highest weight = (0, 0, 0, 1, 0, 0, 0, 0, 0, 0, -3)

Note that power of 11-th root is 2. The roots linear by 11-th
prime root combine into spinorial representation of SO(20), which
does not have vectors with zero weights, so the new Cartan
generator appeared immediately on the first possible level, namely
that of second power of $e_{11}$. This root is imaginary, with
square -2, so it is possible that number of corresponding
generators is greater than one. Of course that will not mean that
we have found more than 11 Cartan generators, since they may not
commute with each other.

One can repeat this calculation for different $E_n$ groups, with
some subtleties for n even.

For the $E_9$ group, with Dynkin diagram

\begin{align}\label{e9}
\begin{picture}(0,100)(0,-50)
\put(-60,0){\line(1,0){140}} \put(-60,0){\circle*{10}}
\put(-40,0){\circle*{10}} \put(-20,0){\circle*{10}}
\put(0,0){\circle*{10}} \put(20,0){\circle*{10}}
\put(40,0){\circle*{10}} \put(40,0){\line(0,1){30}}
\put(40,30){\circle*{10}}
\put(60,0){\circle*{10}}\put(80,0){\circle*{10}}
\put(-62,-15){\text{1}}
\put(-42,-15){\text{2}}\put(-22,-15){\text{3}}
\put(-2,-15){\text{4}}\put(18,-15){\text{5}}
\put(38,-15){\text{6}}\put(58,-15){\text{8}}
\put(78,-15){\text{9}}\put(37,37){\text{7}}
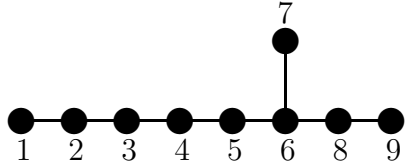
\put(-60,-50){\text{Dynkin diagram of $E_9$ algebra}}
\end{picture}
\end{align}

In the case of $E_9$ if we take out the rightmost root we will be
left with $SO(16)$ diagram which has compact sub-algebra $SO(8)
\times SO(8)$ with rank 8. So to prove that Chevalley-invariant
sub-algebra of $E_9$ has rank 9 we have to find a generator which
commutes with all Cartan generators of $SO(16)$. In analogy with
$E_{11}$ above, we see that there exists a root that has zeros on
first 8 positions of its weight vector, hence commutes with all
Cartan generators of $SO(16)$. That root is found on level 30 (and
actually is the only one on that level) and has the form:

(1, 2, 3, 4, 5, 6, 3, 4, 2)

This proves that compact sub-algebra of $E_9$ has rank 9.

$E_{10}$ has a following Dynkin diagram:

\begin{align}\label{e10}
\begin{picture}(0,100)(0, -50)
\put(-80,0){\line(1,0){160}}\put(-80,0){\circle*{10}}
\put(-60,0){\circle*{10}} \put(-40,0){\circle*{10}}
\put(-20,0){\circle*{10}} \put(0,0){\circle*{10}}
\put(20,0){\circle*{10}} \put(40,0){\circle*{10}}
\put(40,0){\line(0,1){30}} \put(40,30){\circle*{10}}
\put(60,0){\circle*{10}}\put(80,0){\circle*{10}}
\put(-82,-15){\text{1}}
\put(-62,-15){\text{2}}\put(-42,-15){\text{3}}
\put(-22,-15){\text{4}}\put(-2,-15){\text{5}}
\put(18,-15){\text{6}}\put(38,-15){\text{7}}
\put(58,-15){\text{9}}\put(78,-15){\text{10}}
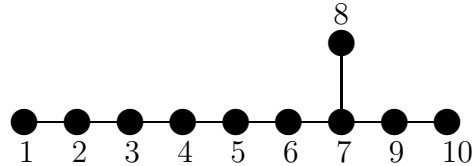
\put(37,37){\text{8}} \put(-80, -50){\text{Dynkin diagram of E10
algebra}}
\end{picture}
\end{align}

For $E_{10}$ we take out the second root from the left (i.e. root
number two), remaining diagram is that of $SO(3)\times E_8$. Its
compact subalgebra is $SO(2)\times SO(16)$ and has the same rank
9. So, as for $E_{11}$, we can change to new invariant Cartan
basis and look for a root with 0 at first and last 8 positions in
its weight vector. Then corresponding generator(s) would commute
with all Cartan generators and the statement that compact
sub-algebra of $E_{10}$ has rank 10 will be proved. Indeed such
root exists on level 61 and has the following coefficients in its
decomposition over simple roots:

(1, 2, 4, 6, 8, 10, 12, 6, 8, 4)

Module of this root is -2, scalar products with simple roots (i.e.
components of its weight) are ( 0, -1, 0, 0, 0, 0, 0, 0, 0, 0).

\section{$K_{n}$ in a new basis}

In a finite-dimensional simple Lie algebras one can consider
different Cartan basises and they all are equivalent (in a complex
field of coefficients), particularly they are connected through
group automorphism. We are not aware of a similar general
statement for Kac-Moody algebras (although see note at p.141 of
\cite{kac1}), and will accept as a hypothesis that for $E_{11}$ we
can change to the Cartan basis, introduced in the previous
Section, and in this new basis it will be described by the same
expressions (\ref{km})-(\ref{5}), where generators e and f are new
generators, constructed from previous ones (obtained by group
automorphism, in the finite-dimensional analogue). The next
question is: what will Chevalley involution (\ref{chev}) look like
in the new basis. Since it commutes with new Cartan generators and
can be simultaneously diagonalized, we can assume that its action
on the new simple generators is multiplication by plus or minus
one. Actually this statement is not correct generally, for any
involution, as is known from finite-dimensional theory - some
automorphisms of Dynkin diagram can be involved, which give rise
to additional lines in Vogan (or Satake) diagrams
\cite{knapp,helg}, namely the lines indicating order 2 orbits of
involution on the set of simple roots. Note that our maximally
compact basis of Cartan generators corresponds to Vogan type
diagrams. But for appearance of such an automorphisms the Dynkin
diagram should have corresponding property, moreover, for
Chevalley involution, at finite dimensions, particularly for $E_7,
E_8, D_{4n}$, i.e. for algebras with maximal rank of $K_n$
subgroup,  Chevalley involution is represented by multiplication
by minus one of all simple generators.  We shall assume the same
behavior for $E_{11}$. This is second crucial assumption of the
present paper.

Evidently, with this assumption, in the new basis, generators of
$K_{11}$ can be described as all even generators of $E_{11}$, and
their roots are all even roots of $E_{11}$.

Now we shall try to construct these roots from some basic even
roots. We would like to introduce the following subalgebra of
$E_{11}$: take the roots of following generators (Lie algebra
commutators are implied)

\begin{eqnarray} \label{ee11g}
a_1= e_7 e_8 e_9 e_{10}, a_2=e_1 e_2,  a_3=e_3 e_4, a_4=e_5 e_6,\\
\nonumber a_5=e_7 e_8, a_6=e_{10} e_{11}, a_7=e_8 e_9, a_8=e_6
e_7,\\ \nonumber a_9=e_8 e_{10},a_{10}=e_4 e_5,  a_{11}=e_2 e_3,\\
\nonumber
\end{eqnarray}

Definition of $a_1$ is actually unique, up to overall sign, since
although Lie brackets can be arranged in different ways, results
coincide. Then one can find corresponding Cartan matrix and Dynkin
diagram:

\begin{eqnarray}\label{o11a}
\begin{array}{*{20}c}
   2&0&0&-1&0&0&0&0&0&0&0 \\
    0&2&-1&0&0&0&0&0&0&0&0 \\
    0&-1&2&-1&0&0&0&0&0&0&0 \\
    -1&0&-1&2&-1&0&0&0&0&0&0 \\
    0&0&0&-1&2&-1&0&0&0&0&0 \\
    0&0&0&0&-1&2&-1&0&0&0&0 \\
    0&0&0&0&0&-1&2&-1&0&0&0 \\
    0&0&0&0&0&0&-1&2&-1&-1&0 \\
    0&0&0&0&0&0&0&-1&2&0&0 \\
    0&0&0&0&0&0&0&-1&0&2&-1 \\
    0&0&0&0&0&0&0&0&0&-1&2
\end{array}
\end{eqnarray}

\begin{align}\label{ee11}
\begin{picture}(0,100)(0,-50)
\put(-80,0){\line(1,0){160}}\put(-80,0){\circle*{10}}
\put(-60,0){\circle*{10}} \put(-40,0){\circle*{10}}
\put(-40,0){\line(0,1){30}} \put(-40,30){\circle*{10}}
\put(-20,0){\circle*{10}} \put(0,0){\circle*{10}}
\put(20,0){\circle*{10}} \put(40,0){\circle*{10}}
\put(40,0){\line(0,1){30}} \put(40,30){\circle*{10}}
\put(60,0){\circle*{10}}\put(80,0){\circle*{10}}
\put(60,0){\circle*{10}}\put(80,0){\circle*{10}}
\put(-82,-15){\text{2}}
\put(-62,-15){\text{3}}\put(-42,-15){\text{4}}
\put(-22,-15){\text{5}}\put(-2,-15){\text{6}}
\put(18,-15){\text{7}}\put(38,-15){\text{8}}
\put(56,-15){\text{10}}\put(78,-15){\text{11}}\put(37,37){\text{9}}
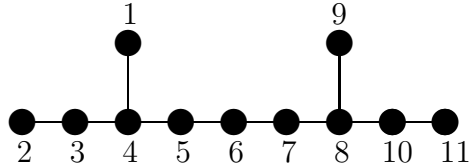
\put(-42,37){\text{1}} \put(-80, -50){\text{Dynkin diagram of
$EE_{11}$ algebra}}
\end{picture}
\end{align}

where simple roots in (\ref{ee11}) are enumerated in agreement
with (\ref{o11a}). One can construct the corresponding abstract
Kac-Moody algebra, we denote it by $EE_{11}$ since it contains two
E type tails, and this notation is similar to that of hyperbolic
algebras - AE, BE, CE, DE. All roots (\ref{ee11g}) are real, hence
according to \cite{n1}, algebra (\ref{ee11}) is isomorphic to
subalgebra in $E_{11}$, generated by generators (\ref{ee11g}). Our
hypothesis is that perhaps this algebra has the same roots as
$K_{11}$, i.e. all even roots of $E_{11}$. Statement seems to be
simple, nevertheless, we were not able to prove it algebraically,
due to unknown structure of roots system. Instead we checked that
up to level 146 by the help of computer program, which generates
the roots for an arbitrary input Dynkin diagram. The number of
roots up to the level 146 (inclusively) is 19661788 (without
counting multiplicity), so coincidence is considerable. Since
multiplicity of real roots is one, this statement means that these
two algebras coincide at least in a sector of real roots. It is
not clear whether such coincidence exists also for imaginary roots
(that would mean that groups $K_{11}$ and $EE_{11}$ coincide), at
least imaginary root generators of $EE_{11}$ belong to $K_{11}$.

The similar statements are correct for $E_{10}$ and $E_{9}$
groups. Corresponding composite roots (generators) and Dynkin
diagram for $E_{10}$ are:

\begin{eqnarray} \label{de10-1}
a_1=e_6 e_7 e_8 e_9, a_2=e_2 e_3, a_3=e_4 e_5, a_4=e_6 e_7,
a_5=e_9 e_{10}, \\a_6=e_7 e_8, a_7=e_5 e_6, a_8=e_7 e_9, a_9=e_3
e_4, a_{10}=e_1 e_2\nonumber
\end{eqnarray}

\begin{align}\label{de10}
\begin{picture}(0,100)(0,-50)
\put(-80,0){\line(1,0){140}}\put(-80,0){\circle*{10}}
\put(-60,0){\circle*{10}} \put(-40,0){\circle*{10}}
\put(-60,0){\line(0,1){30}} \put(-60,30){\circle*{10}}
\put(-20,0){\circle*{10}} \put(0,0){\circle*{10}}
\put(20,0){\circle*{10}} \put(40,0){\circle*{10}}
\put(20,0){\line(0,1){30}} \put(20,30){\circle*{10}}
\put(60,0){\circle*{10}} \put(-82,-15){\text{2}}
\put(-62,-15){\text{3}}\put(-42,-15){\text{4}}
\put(-22,-15){\text{5}}\put(-2,-15){\text{6}}
\put(18,-15){\text{7}}\put(38,-15){\text{9}}
\put(58,-15){\text{10}}\put(17,37){\text{8}}
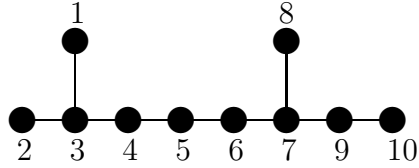
\put(-62,37){\text{1}} \put(-80, -50){\text{Dynkin diagram of
$DE_{10}$ algebra}}
\end{picture}
\end{align}
\vspace{5mm}

For $E_{10}$ we check our hypothesis up to level 200, the number
of roots up to that level is 18441198, and no deviation is found.

For $E_{9}$:

\begin{eqnarray} \label{d91c}
a_1=e_1 e_2, a_2=e_5 e_6 e_7 e_8, a_3=e_3 e_4, a_4=e_5 e_6,\\
a_5=e_8 e_9, a_6=e_6 e_7, a_7=e_4 e_5, a_8=e_6 e_8, a_9=e_2 e_3
\nonumber
\end{eqnarray}

\begin{align} \label{d91}
\begin{picture}(0,100)(0,-50)
\put(-60,0){\line(1,0){120}} \put(-60,0){\circle*{10}}
\put(-40,0){\circle*{10}} \put(-40,0){\line(0,1){30}}
\put(-40,30){\circle*{10}} \put(-20,0){\circle*{10}}
\put(0,0){\circle*{10}} \put(20,0){\circle*{10}}
\put(40,0){\circle*{10}} \put(40,0){\line(0,1){30}}
\put(40,30){\circle*{10}} \put(60,0){\circle*{10}}
\put(-62,-15){\text{2}}\put(-42,-15){\text{3}}
\put(-22,-15){\text{4}}\put(-2,-15){\text{5}}
\put(18,-15){\text{6}}\put(38,-15){\text{7}}
\put(58,-15){\text{9}}\put(37,37){\text{8}} \put(-42,37){\text{1}}
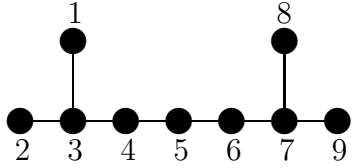
\put(-80, -50){\text{Dynkin diagram of $D_9^{(1)}$ algebra}}
\end{picture}
\end{align}

This affine $E_9$ case is easy to check by computer, due to linear
(not exponential, as in other, non-affine algebras) grow of number
of roots. We check situation up to the level 10000, with complete
confirmation of coincidence of roots of $D_9^{(1)}$ and even roots
of $E_9$. Here we can make a conclusion on complete coincidence,
including imaginary roots, since both algebras - $D_9^{(1)}$ and
$E_9$ are affine Kac-Moody algebras, multiplicity of imaginary
roots of which depends on the rank only, hence not only even
roots, but even generators of $E_9$ coincide with all generators
of $D_9^{(1)}$. This conclusion disagrees with the statement of
\cite{n6}. It may mean that some of our assumptions are incorrect:
either a possibility of changing to the new, invariant Cartan
basis, or representation of Chevalley involution in new basis as
reflection of all simple generators.

One can provide an algebraic proof that all even generators of
$E_9$ coincide with all generators of $D_9^{(1)}$. This is due to
the fact, that the set of all roots of the (untwisted) affine
algebra is completely known (see convenient physical exposition in
\cite{godd}). Namely, that is the set of all non-zero triplets
$(\alpha, 0, n)$, where $\alpha$ are all roots of finite algebra,
on which affine one is based, including zero. Positive roots are
those with $n>0$, and, at $n=0$, roots $(\alpha_+,0,0)$, with
positive roots at first place. The set of simple roots is
$(\alpha_i, 0,0)$ (corresponding to the roots of finite algebra)
and (additional affine root) $(-\psi,0,1)$, where $\psi$ is
maximal positive root of finite algebra. The set of even roots is
$(\alpha, 0, n)$ with even $\alpha$. We can take the set of all
positive roots of $D_9^{(1)}$, substitute the expressions
(\ref{d91c}) of roots of $D_9^{(1)}$ through the roots of $E_9$
and check that we obtain in this way all positive even roots of
$E_9$. Denote all roots of $D_8$ by $\beta=\sum\limits_{k = 2}^9
{m_k \beta_k}$ where $\beta_k, k=2\div 9$ are simple roots, and
set ${m_k}$ here and below exactly parameterize all roots of
$D_8$. We denote simple affine roots $\beta_1$ for $D_9^{(1)}$ and
$\alpha_1$ for $E_9$ and use the same notation $\beta_k, k=2\div
9$ for remaining simple roots $(\beta_i, 0,0)$ of $D_9^{(1)}$. So,
expanding an arbitrary $D_9^{(1)}$ root over simple roots, we
have:

\begin{eqnarray} \label{x}
(\beta, 0,n)=(\sum\limits_{k = 2}^9 {m_k
\beta_k},0,n)=\sum\limits_{k = 1}^9 {n_k \beta_k}\\
n_1=n, n_k=n n_k^{\psi}+m_k
\end{eqnarray}
where $n_k^{\psi}$ are coefficients of expansion of highest root:
\begin{equation}
\psi=\sum\limits_{k = 2}^9 {n_k^{\psi} \beta_k}
\end{equation}

Substituting expressions (\ref{d91c}) for $D_9^{(1)}$ roots
$\beta$ through roots $\alpha$ of $E_9$, we obtain
\begin{eqnarray} \label{xx}
(\beta, 0,n)=(n(-\lambda +\alpha_2)+ \sum\limits_{k = 2}^9 {n
n_k^{\psi} \beta_k(\alpha)} + \sum\limits_{k = 2}^9 {m_k
\beta_k(\alpha)},0,n)
\end{eqnarray}
where $\lambda$ is highest positive root of $E_8$. The sum of
first two terms in (\ref{xx}) is zero:
\begin{eqnarray} \label{xxx}
n(-\lambda +\alpha_2)+\sum\limits_{k = 2}^9 {n n_k^{\psi}
\beta_k(\alpha)}
\end{eqnarray}
which can be easily checked with the help of (\ref{d91c}) and
components of highest roots:
\begin{eqnarray} \label{xx4}
\psi \sim (12222211)\\
\lambda \sim (23456342)
\end{eqnarray}
Final form of equation (\ref{xx})
\begin{eqnarray} \label{x5}
(\beta, 0,n)= (\sum\limits_{k = 2}^9 {m_k \beta_k(\alpha)},0,n)
\end{eqnarray}
is exactly the statement that even levels of $E_8$ coincide with
group SO(16). This statement is the finite dimensional analog and
basis of our statements on Kac-Moody algebras $E_n, n=11,10,9$.

\section{On a representations of $EE_{11}$}

One can easily obtain the weights of the highest weight vectors of
fundamental representations of $E_{11}$. These are the rows of an
inverse Cartan matrix:

\begin{eqnarray}\label{oinv}
EE_{11}^{-1}=\frac{1}{8} \left[ \begin{array}{*{20}c}
-1&2&4&6&9&12&15&18&9&12&6 \\
    2&-4&0&4&6&8&10&12&6&8&4 \\
    4&0&0&8&12&16&20&24&12&16&8 \\
    6&4&8&12&18&24&30&36&18&24&12 \\
    9&6&12&18&15&20&25&30&15&20&10 \\
    12&8&16&24&20&16&20&24&12&16&8 \\
    15&10&20&30&25&20&15&18&9&12&6 \\
    18&12&24&36&30&24&18&12&6&8&4 \\
    9&6&12&18&15&12&9&6&-1&4&2 \\
    12&8&16&24&20&16&12&8&4&0&0 \\
    6&4&8&12&10&8&6&4&2&0&-4
    \end{array} \right]
\end{eqnarray}

We would like to compare these weights with those of $E_{11}$:

\begin{eqnarray}\label{einv}
E_{11}^{-1}=\frac{1}{2} \left[ \begin{array}{*{20}c}
-1&0&1&2&3&4&5&6&3&4&2 \\
    0&0&2&4&6&8&10&12&6&8&4 \\
    1&2&3&6&9&12&15&18&9&12&6 \\
    2&4&6&8&12&16&20&24&12&16&8 \\
    3&6&9&12&15&20&25&30&15&20&10 \\
    4&8&12&16&20&24&30&36&18&24&12 \\
    5&10&15&20&25&30&35&42&21&28&14 \\
    6&12&18&24&30&36&42&48&24&32&16 \\
    3&6&9&12&15&18&21&24&11&16&8 \\
    4&8&12&16&20&24&28&32&16&20&10 \\
    2&4&6&8&10&12&14&16&8&10&4
\end{array} \right]
\end{eqnarray}

The only subtlety is that rows of inverse Cartan matrix express
weights in a basis of simple roots of a given algebra, so for
comparison we should express both in the same basis, e.g. in a
basis of simple roots of $E_{11}$. The expression of simple roots
of $EE_{11}$ through the simple roots of $E_{11}$ is given in
(\ref{ee11g}). So we should multiply the matrix (\ref{oinv}) from
the right by the transformation matrix

\begin{eqnarray}\label{t}
T=\left[ \begin{array}{*{20}c}
0&0&0&0&0&0&1&1&1&1&0 \\
    1&1&0&0&0&0&0&0&0&0&0 \\
    0&0&1&1&0&0&0&0&0&0&0 \\
    0&0&0&0&1&1&0&0&0&0&0 \\
    0&0&0&0&0&0&1&1&0&0&0 \\
    0&0&0&0&0&0&0&0&0&1&1 \\
    0&0&0&0&0&0&0&1&1&0&0 \\
    0&0&0&0&0&1&1&0&0&0&0 \\
    0&0&0&0&0&0&0&1&0&1&0 \\
    0&0&0&1&1&0&0&0&0&0&0 \\
    0&1&1&0&0&0&0&0&0&0&0
\end{array} \right]
\end{eqnarray}

and obtain

\begin{eqnarray}\label{oinvt}
(EE_{11})^{-1}T=\frac{1}{8} \left[ \begin{array}{*{20}c}
2&8&10&16&18&24&26&32&14&20&12 \\
    -4&0&4&8&12&16&20&24&12&16&8 \\
    0&8&8&16&24&32&40&48&24&32&16 \\
    4&16&20&32&36&48&60&72&36&48&24 \\
6&16&22&32&38&48&54&64&34&44&20\\
      8&16&24&32&40&48&56&64&32&40&16\\
      10&16&26&32&42&48&58&64&30&44&20\\
     12&16&28&32&44&48&60&72&36&48&24\\
      6&8&14&16&22&24&30&32&18&20&12\\
      8&8&16&16&24&32&40&48&24&32&16\\
      4&0&4&8&12&16&20&24&12&16&8\\
\end{array} \right]
\end{eqnarray}

The second row of (\ref{oinvt}) and first row of (\ref{einv})
coincide. This is a very interesting coincidence, since exactly
the first fundamental representation of $E_{11}$ is shown, in
\cite{w1}, to contain the brane charges of 11d supergravity. The
brane charges are appearing on the right hand side of the
supersymmetry relation:
\begin{eqnarray}\label{11d}
\left\{ \bar{Q},Q\right\} &=&\Gamma ^{\mu}P_{\nu}+\Gamma
^{\mu\nu}Z_{\mu\nu}+\Gamma ^{\mu\nu\lambda\rho\sigma}
Z_{\mu\nu\lambda\rho\sigma}, \label{0}\\ \mu, \nu,...
&=&0,1,2,..10. \nonumber
\end{eqnarray}

This coincidence resembles the statement of \cite{wit}, where it
is noticed that at certain dimensions certain brane charges are
not only in the representation of corresponding susy algebra's
automorphism group, but also combine into representations of
U-duality group. For example, consider maximal susy algebra in 4
dimensions, obtained from reduction to 4d of 11d algebra
(\ref{11d}). The corresponding 4d algebra has  SU(8) as an
automorphism group. Scalar central charges appear from few
sources: 7 from vector, 21 from membrane charge, and 28 from
five-brane charge, altogether they combine into 56 of 4d U-duality
group $E_7$, which includes SU(8) as its maximal compact subgroup.
This and other similar statements of \cite{wit} is interesting to
compare with above statement of coincidence of weights of $E_{11}$
and $EE_{11}$, this is under investigation. Particularly, such a
coincidence can reflect the possibility of extension of
corresponding representation of $EE_{11}$ to the $l_1$
representation of $E_{11}$.

\section{Conclusion}

In the present paper we consider the $E_{n}$ and $K_{n}$groups,
involved in different symmetry hypothesis of M-theory,
particularly non-linear realization, at n=11, 10, 9. We make few
hypothesis, extending some facts from finite group theory to
Kac-Moody case. These crucial assumptions are first, possibility
to change to new basis of Cartan generators, and, second,
representation of Chevalley involution as reflection
(multiplication by (-1)) of all simple roots generators. Both
statements are supported in finite-dimensional simple Lie
algebras, but are not clear for Kac-Moody algebras. With this
assumption we present an observation, based on computer check up
to some high levels of roots, that for $E_{11}$, $E_{10}$ and
$E_{9}$ algebras roots of compact (Chevalley-invariant) subalgebra
coincide with roots of some Kac-Moody algebra, corresponding
Dynkin diagrams are given by (\ref{ee11}), (\ref{de10}) and
(\ref{d91}) correspondingly.

As noticed in body of paper, above result means that generators,
corresponding to real roots, coincide, since they are unique. For
further study of hypothesis mentioned, one has to calculate the
multiplicities of imaginary roots and compare them.

The naturally arising problem is to reformulate in a new basis the
known properties of E/K nonlinear model, e.g. the supergravity
solutions. In \cite{west} they have been represented purely in the
group language, so it seems to be a reasonable task. It may appear
that some new solutions will naturally appear in a new basis.
Another important question is to find an image of $A_{10}$
subgroup, which serves in \cite{w0,w2,w1} to purpose of finding a
contact with usual formulations of supergravities.

Besides,  the new basis has an advantage that it naturally leads
to the subgroups,  which belong to compact subgroup $K_{11}$, and,
particularly, naturally extend Lorentz group (which appears as a
compact subgroup of $A_{10}$, roots 1-10 at (\ref{e11}). Different
subgroups of $K_{11}$ can be read off directly from Dynkin diagram
(\ref{ee11}) of its subgroup $EE_{11)}$. Finally, it is worth to
mention again the problem of finding a relations between $l_1$
representation of $E_{11}$ and that of $EE_{11}$.

\section{Acknowledgements}

We are indebted to R.Manvelyan for discussions. Work is partly
supported by INTAS grant 03-51-6346.

\end{document}